\documentclass[journal]{IEEEtran}
\IEEEoverridecommandlockouts
\usepackage{cite}
\usepackage{amsmath,amssymb,amsfonts}
\usepackage{amsthm,bm}
\usepackage{graphicx}
\usepackage{textcomp}
\usepackage{url}
\usepackage{mathtools}
\def\BibTeX{{\rm B\kern-.05em{\sc i\kern-.025em b}\kern-.08em
    T\kern-.1667em\lower.7ex\hbox{E}\kern-.125emX}}

\usepackage{stackengine}
\usepackage{algorithm}
\usepackage{booktabs}
\usepackage{subfig}
\usepackage{multicol}
\usepackage[noend]{algpseudocode}
\usepackage[table,xcdraw]{xcolor}
\makeatletter
\def\BState{\State\hskip-\ALG@thistlm}
\makeatother

\algnewcommand\algorithmicforeach{\textbf{for each}}
\algdef{S}[FOR]{ForEach}[1]{\algorithmicforeach\ #1\ \algorithmicdo}
    
\begin{document}

\title{\huge Realizing the Metaverse with Edge Intelligence: \\
A Match Made in Heaven}

\author{Wei Yang Bryan Lim\thanks{WYB.~Lim is with Alibaba Group and Alibaba-NTU Joint Research Institute (JRI), Nanyang Technological University (NTU), Singapore. Email: limw0201@e.ntu.edu.sg. Z.~Xiong is with Singapore University of Technology and Design, Information Systems Technology and Design (ISTD) Pillar, Singapore. Email: zehui\_xiong@sutd.edu.sg. D.~Niyato and C.~Miao are with School of Computer Science and Engineering, NTU, Singapore. Emails: dniyato@ntu.edu.sg and ascymiao@ntu.edu.sg. X.~Cao is with School of Electronic and Information Engineering, Beihang University, Beijing, China. Email: xbcao@buaa.edu.cn. S.~Sun is with Communications and Networks Department, Institute for Infocomm Research (I2R), Singapore. Email: sunsm@i2r.a-star.edu.sg. Q. Yang is with Department of Computer Science and Engineering, Hong Kong University of Science and Technology, Hong Kong, and Webank. Email: qyang@cse.ust.hk.}, Zehui Xiong, Dusit Niyato, \textit{IEEE Fellow}, Xianbin Cao, Chunyan Miao, \\ Sumei Sun, \textit{IEEE Fellow}, Qiang Yang, \textit{IEEE Fellow}}

\makeatletter
\setlength{\@fptop}{0pt}
\makeatother

\maketitle

\begin{abstract}

Dubbed ``the successor to the mobile Internet", the concept of the Metaverse has recently exploded in popularity. While there exists lite versions of the Metaverse today, we are still far from realizing the vision of a seamless, shardless, and interoperable Metaverse given the stringent sensing, communication, and computation requirements. Moreover, the birth of the Metaverse comes amid growing privacy concerns among users. In this article, we begin by providing a preliminary definition of the Metaverse. We discuss the architecture of the Metaverse and mainly focus on motivating the convergence of edge intelligence and the infrastructure layer of the Metaverse.  We present major edge-based technological developments and their integration to support the Metaverse engine. Then, we present our research attempts through a case study of virtual city development in the Metaverse. Finally, we discuss the open research issues.

\end{abstract}


\newtheorem{definition}{Definition}
\newtheorem{lemma}{Lemma}
\newtheorem{theorem}{Theorem}

\newtheorem{property}{Property}

\section{Introduction}
\label{sec:intro}

The concept of \textit{Metaverse} first appeared in the science fiction novel \textit{Snow Crash} written by Neal Stephenson in 1992. More than twenty years later, the Metaverse has re-emerged as a buzzword. In short, the Metaverse is commonly described as an embodied version of the Internet. Just as how we navigate today's Internet with a mouse cursor, users will explore the Metaverse with the aid of virtual reality (VR) or augmented reality (AR) technologies. Moreover, powered by Artificial Intelligence (AI), blockchain technology, and 5G and Beyond (B5G),  the Metaverse is envisioned to facilitate peer-to-peer interactions and support novel, decentralized ecosystems of service provisions that will blur the lines between the physical and virtual worlds. 

To date, tech giants have invested heavily towards realizing the Metaverse as ``the successor to the mobile Internet". Among others, Facebook was even rebranded as ``Meta" to reinforce its commitment towards the development of the Metaverse. There are two fundamental driving forces behind the excitement surrounding the Metaverse. First, the Covid-19 pandemic has resulted in a paradigm shift in how social interactions are conducted today, thereby positioning the Metaverse as a \textit{necessity} in the near future. Second, emerging technological enablers have made the Metaverse a growing \textit{possibility}. For example, advances in VR/AR and haptic technologies  enable users to be visually and physically immersed in  a virtual world. To date, there exist ``lite" versions of the Metaverse that have evolved mainly from Massive Multiplayer Online (MMO) games. Among others, Roblox\footnote{https://www.roblox.com/} and Fornite\footnote{https://www.epicgames.com/fortnite/en-US/home} started  as online gaming platforms. Yet, just recently, the virtual concerts held on Roblox and Fornite attracted millions of views. 

However, we are still far from realizing the Metaverse. For one, the aforementioned ``lite" versions are distinct platforms operated by separate entities. In other words, one's Fortnite avatar and virtual items mean nothing in the Roblox world. In contrast, the Metaverse is envisioned to be a seamless integration of virtual worlds. Next, while MMO games can host more than a hundred players at once, albeit with high-specification system requirements, an open-world  \textit{VRMMO}  application is still a relatively nascent concept even in the gaming industry. Similarly, it will be a challenge to develop a ``shardless" Metaverse that is persistent, rather than one that separates players into different sessions. This is  exacerbated by the expectation that large parts of the Metaverse have to integrate the physical and virtual worlds, e.g., through digital twins. The stringent sensing, communication, and computation requirements impede the real-time, scalable, and ubiquitous implementation of the Metaverse. Finally, the birth of the Metaverse comes amid increasingly stringent privacy regulations. 

In this article, we begin by motivating a definition and introduction to the architecture of the Metaverse. To realize the Metaverse amid its unique challenges, we mainly focus on the \textit{edge intelligence} driven infrastructure layer,  which is a core feature in B5G wireless networks. In short, edge intelligence is the convergence between edge computing and AI. We adopt the two commonly-quoted  divisions of edge intelligence, i.e., i) \textit{Edge for AI}: which refers to the end-to-end framework of bringing sensing, communication, AI model training, and inference closer to  where data is produced, and ii) \textit{AI for Edge}: which refers to the use of AI algorithms to improve the orchestration of the aforementioned framework. Then, as a case study, we present a framework for the collaborative edge-driven virtual city development in the Metaverse. Finally, we discuss the open research issues.

Our contributions are as follows:
\begin{enumerate}
    \item We present a general architecture of the Metaverse and its major components, thereby providing a holistic view of the Metaverse ecosystems. We outline the key technologies that enable the edge-driven Metaverse, emphasizing their roles to support virtual services.

     \item We discuss potential applications and services that can be delivered in the Metaverse, and through a case study on virtual city development, demonstrate the convergence between edge intelligence and the Metaverse engine.

     \item We present research perspectives and highlight the interdisciplinary open issues and research opportunities.

\end{enumerate}

\section{The Metaverse: Architecture, Technologies, and Applications}

\begin{figure*}[t]
    \centering
    \includegraphics[clip, height= 11cm]{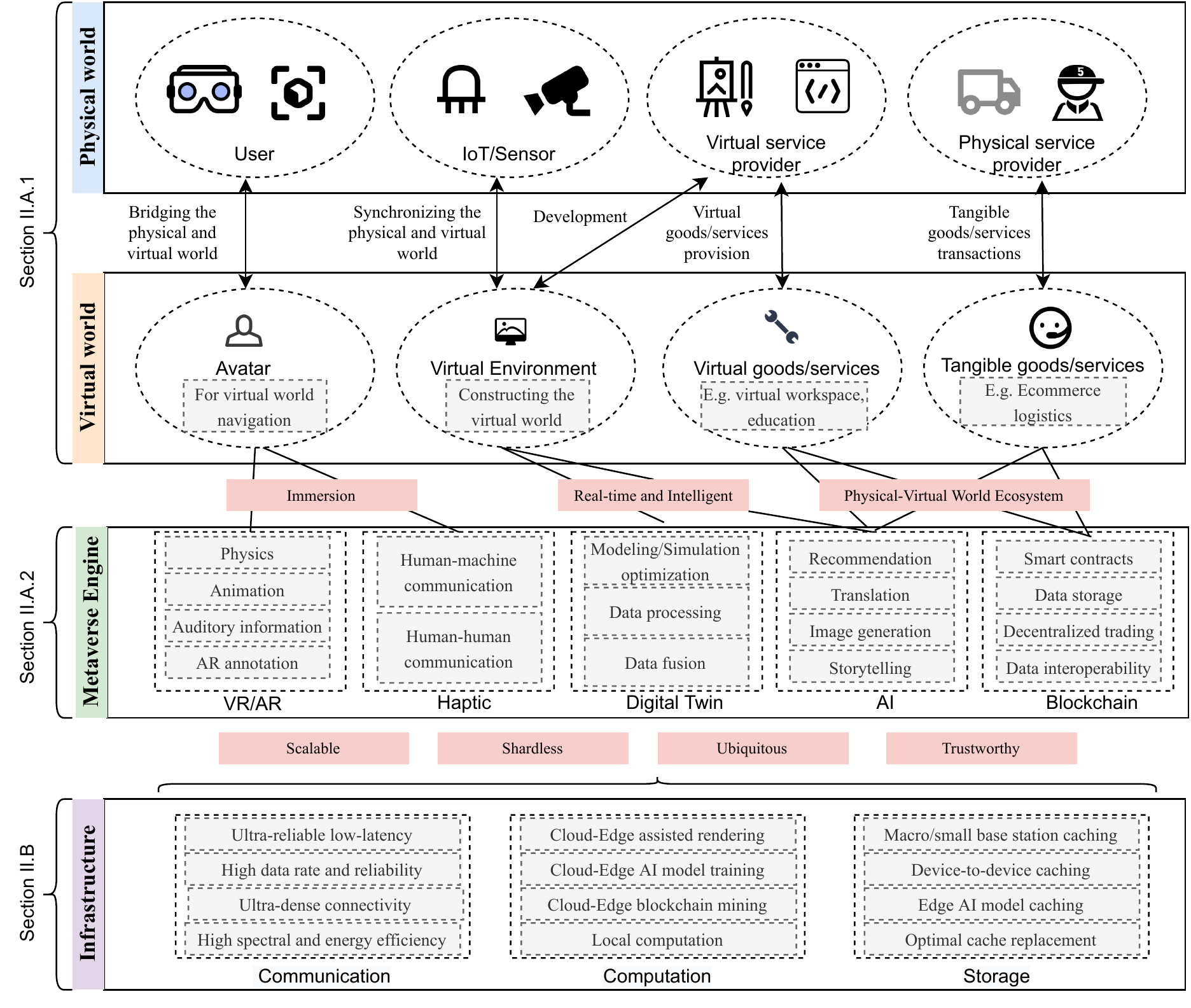}\par 
    \caption{The Metaverse architecture features the immersive and real-time physical-virtual world interaction supported by the Metaverse engine. The supporting infrastructure ensures that the Metaverse is scalable, shardless, enables ubiquitous access, and is trustworthy for users.}
    \label{fig:archi}
\end{figure*}

The Metaverse is an \textit{embodied} version of the Internet that comprises a \textit{seamless} integration of \textit{interoperable, immersive, and shardless virtual ecosystems} navigable by user-controlled avatars. In this section, we present the layers of the Metaverse architecture (Fig. \ref{fig:archi}). 

\subsection{Physical-virtual world and the Metaverse engine}
\label{pvw}

\begin{enumerate}
    
    \item \textit{Physical-virtual world interaction}: Each non-mutually-exclusive stakeholder in the physical world controls components that influence the virtual world. The consequences in the virtual world in turn feedbacks to the physical world. The key stakeholders are:
    
    \begin{itemize}
    
        \item \textit{Users} can experience the virtual world through Head Mounted Displays (HMDs) or AR goggles. The users can in turn execute actions to interact with other users or virtual objects.

        \item \textit{IoT and sensor networks} deployed in the physical world collect data from the  environment. The insights derived are used to update the virtual environment, e.g, through feeding information to update a digital twin. The sensor network may be independently owned by sensing service providers (SSPs) that contribute live data feeds to virtual service providers (VSPs) to generate and maintain the virtual environment.
        
        \item \textit{Virtual service providers (VSPs)} develop and maintain the virtual worlds of the Metaverse. Similar to user-created videos today (e.g., YouTube), the Metaverse is envisioned to be enriched with user-generated content (UGC) that includes virtual art, games, and social applications. These UGC can be traded in the Metaverse.
        
        \item \textit{Physical service providers} operate the physical infrastructure that supports the Metaverse engine and respond to transaction requests that originate from the Metaverse. This includes the operations of communication and computation resources at the edge of the network, or logistics services for the delivery of physical goods transacted in the Metaverse.

    \end{itemize}

    \item The \textit{Metaverse engine} obtains inputs such as data from stakeholder-controlled components. The virtual world is generated, maintained, and enhanced with these inputs.

\begin{figure*}
    \centering
    \includegraphics[clip, height= 7cm]{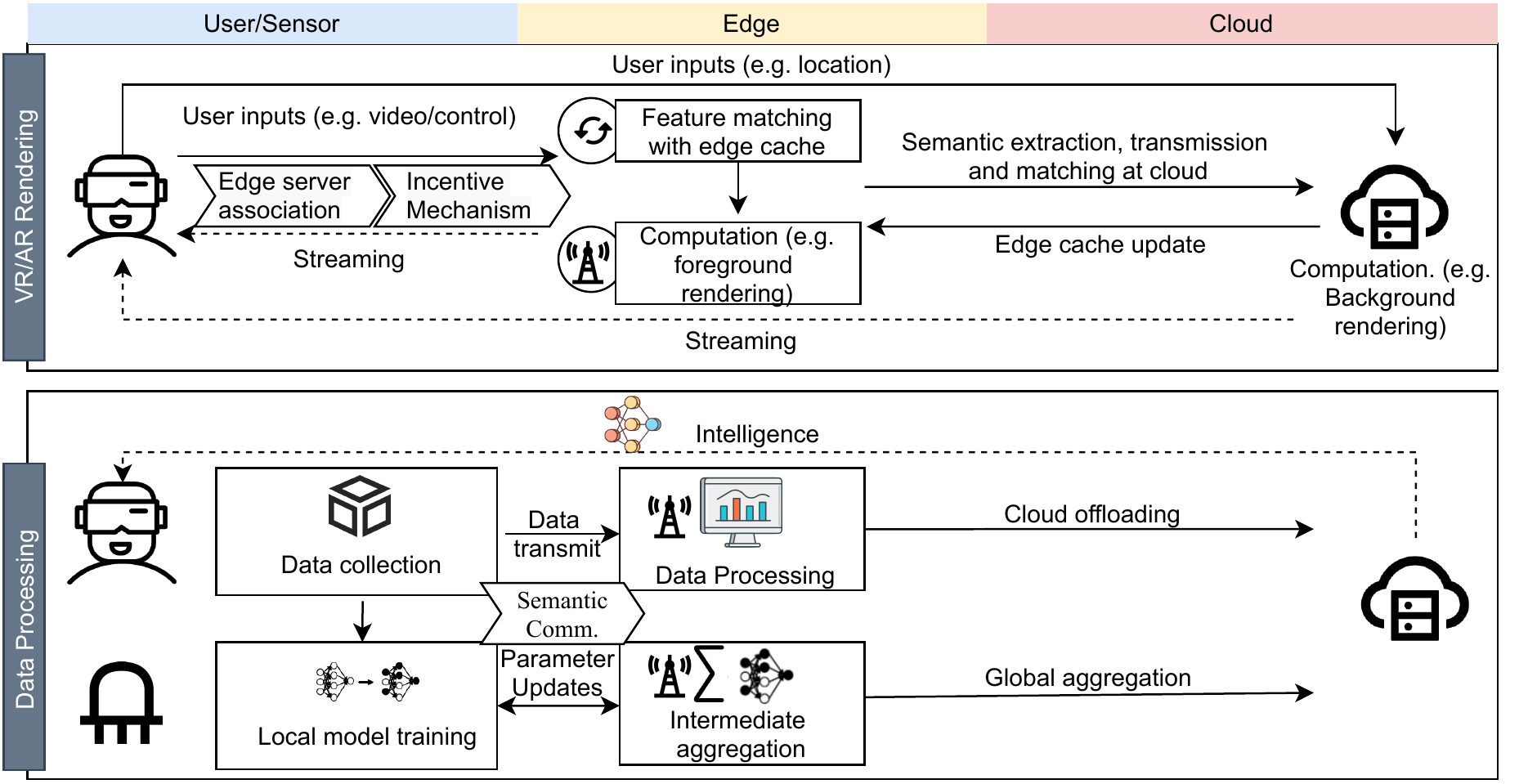}\par 
    \caption{Applications of Edge Intelligence for the Metaverse.}
    \label{fig:ei}
\end{figure*}

    \begin{itemize}
        
        \item \textit{VR/AR} enables users to experience the Metaverse visually, whereas \textit{haptics} enable users to experience the Metaverse through the additional dimension of touch, e.g., using haptic gloves. This enhances user interactions, e.g., through transmitting a handshake across the world, and opens up the possibilities of providing physical services in the Metaverse, e.g., remote surgery. These technologies are developed by standards that facilitate interoperability, e.g., Virtual Reality Modelling Language (VRML)\footnote{\url{https://www.web3d.org/documents/specifications/14772/V2.0/part1/javascript.html}}, that govern the properties, physics, animation, and rendering of virtual assets, so that users can traverse the Metaverse smoothly.

        \item \textit{Digital twins} enable some  virtual worlds within the Metaverse to be modeled after the physical world in real-time. This is accomplished through modeling and data fusion. Digital twins add to the realism of the Metaverse and facilitates new dimensions of services and social interaction. For example, Microsoft Mesh  allows users working from multiple sites to collaborate with each other in real-time digital copies of their office.
        
        \item \textit{Artificial Intelligence} can be leveraged to incorporate intelligence into the Metaverse for improved user experience, e.g., for efficient object rendering, intelligent chatbots, and UGC. For example, the MetaHuman project\footnote{\url{https://www.unrealengine.com/en-US/digital-humans}} by EpicGames utilizes AI to generate life-like digital characters quickly. The generated characters may be deployed by VSPs as conversational virtual assistants to populate the Metaverse.
        
        \item \textit{Blockchain} technology will be key to preserving the value and universality of virtual goods, as well as establishing the economic ecosystem within the Metaverse. It is difficult for current virtual goods to be of value outside the platforms on which they are traded or created. Blockchain technology will play an essential role in reducing the reliance on such centralization. For example, a Non-fungible token (NFT) serves as a mark of a virtual good's uniqueness and authenticates one's ownership of the good. This protects the value of virtual goods and facilitates the peer-to-peer trading in a decentralized environment. As virtual worlds in the Metaverse are developed by different parties, the user data may also be managed separately. To enable seamless traversal across  virtual worlds, multiple parties will need to access and operate on such user data. Due to value isolation among blockchains,  cross-chain is a crucial technology to enable secure data interoperability.

    \end{itemize}
    
    % The Metaverse engine provides users with an immersive experience. Using data from SSPs, parts of the virtual world can be developed by VSP to reflect the physical world in real-time. The  stream of data from users and SSPs fed into the Metaverse engine is essential to maintain the intelligent physical-virtual world ecosystem.
    
    \subsection{Edge intelligence-empowered infrastructure}

     The general functions of the infrastructure layer are:
    
    \begin{itemize}
    
    \item \textit{Communication and Networking:} To prevent breaks in presences (BIP), i.e., disruptions that cause a user to be aware of the real world setting, VR requires a data rate of $250$ Mbit/s and packet error rate on the order of $10^{-1} \sim 10^{-3}$. Haptic traffic requires a lower data rate of $1$ Mbit/s and  packet error rate on the order of $10^{-4} \sim 10^{-5}$ \cite{park2018urllc}. This may be enabled through enhanced mobile broadband (eMBB) and ultra-reliable and low latency communication (URLLC) links, which are the main techonlogy pillars in B5G.  Due to the expected explosive growth of data traffic, ultra-dense networks  deployed in B5G networks to alleviate the constrained system capacity. 
    
    \item \textit{Computation and Storage:} Today, MMO games can host more than a hundred players in a single game session and hence require high-specification GPU requirements. VRMMO games, which are the rudiment of the Metaverse system, are still scarce in the industry and may require the devices such as HMDs to be connected to powerful computers to render both the immersive virtual worlds and the interactions with hundreds of other players.  To enable ubiquitous access to the Metaverse, a promising solution is the cloud-edge-end computation paradigm. Specifically, local computations  can be performed on end devices for the least resource consuming task, e.g., computations required by the physics engine to determine the movement and positioning of an avatar. To reduce the burden on the cloud for scalability, and further reduce end-to-end latency, edge servers can be leveraged to perform costly foreground rendering, which requires less graphical details but lower latency \cite{guo2020adaptive}. The more computation intensive but less delay sensitive tasks, e.g., background rendering, can in turn be executed on cloud servers. Moreover, popular contents can be cached at the edge of the network for efficient retrieval and reduction in computation overheads.
    
    \end{itemize}
    
    The infrastructure layer leverages edge intelligence (Fig. \ref{fig:ei}) to (i) support AI for the intelligent Metaverse (i.e., Edge for AI), and (ii) utilize AI to realize the resource-efficient collaborative edge paradigm (i.e., AI for Edge). 
    
    \begin{itemize}
        
        \item \textit{Edge for AI}

            \textit{Edge offloading:}  Apart from offloading rendering computations to the edge or cloud, costly computation tasks required for data processing and AI model training, e.g., matrix multiplication, can also be decomposed into subtasks to be offloaded to edge servers (i.e., workers). The completed subtasks are aggregated at a master node to recover the computation result. However, a major drawback of computation offloading is the existence of stragglers, which are the processing nodes that run slower than expected or nodes that may be disconnected from the network due to several factors such as imbalanced work allocation and network congestion. As a result, the overall time needed to execute the task is determined by the slowest processing node. One way to mitigate the straggler effect is to utilize worker selection schemes to eliminate straggling workers. Another way is to leverage coded redundancy to reduce the recovery threshold, i.e., the number of workers that need to submit their results for the master to reconstruct the final result. For example, polynomial codes \cite{yu2017polynomial} can be used to generate redundant intermediate computations. The total computation is not determined by the slowest straggler but by the time taken for the master node to receive computed results from some decodable set of workers. For polynomial codes, the recovery threshold does not scale with the number of workers involved, thereby ensuring the scalability of the edge-empowered Metaverse.
            
             \textit{Caching:} Edge caching is instrumental to reduce computation and communication redundancy, which refers to the wastage of network resources as a result of repetitive user access of popular content or computations. For the former, the probabilistic model for the popularity distribution of files, e.g., field of views (FOV) in the Metaverse, can be learned \cite{sun2019communications}. Then, the popular FOVs can be stored at edge servers close to users that demand it more to reduce rendering computation cost and latency. For the latter, the computation results from AI models can be cached at edge servers to respond to computation requests that are of a similar nature. Moreover, pre-trained models can be cached at the edge to perform costly inference tasks for faster response to users.
            
             \textit{Local machine learning model training:} As with the Internet, the Quality of Experience (QoE) that users derive from the Metaverse  will  improve with more insights gathered from usage data. However, following the introduction of increasingly stringent privacy laws such as the General Data Protection Regulation (GDPR), the Metaverse will have to be developed while preserving user privacy. Moreover, the risk of data leaks increases in tandem with the increase in attack surfaces as more users are connected to the Metaverse. One  solution is the privacy-preserving machine learning paradigm known as Federated Learning (FL) \cite{mcmahan2017communication}. In FL, users of the Metaverse can carry out AI model training on their local device before transmitting the model parameters or gradient updates, instead of the raw data, to the model owner for aggregation. This enables privacy-preserving collaborative machine learning while leveraging the computation capabilities of these users, e.g., during idle device usage periods. As model parameters are smaller in size than raw data, FL also alleviates the burden on backbone communication networks. Recently, the edge-assisted Hierarchical FL framework  have also been proposed \cite{abad2020hierarchical} in which intermediate model aggregations are performed at edge servers before global cloud aggregation, so as to reduce link distances and instances of costly global communication with the cloud.

        \item \textit{AI for Edge}

            \textit{Semantic communication:} The advent of the Metaverse will inevitably contribute to a growing demand for bandwidth amid the explosive data traffic volume required to support the Metaverse engine. This necessitates a paradigm shift from Shannon's conventional focus in how accurately the communication symbols can be transmitted to how precisely the transmitted symbols can convey the \textit{meaning} of the message. In particular, the human-to-machine (H2M) semantic communication can be a key technology to optimize VR/AR implementation for the ubiquitous Metaverse \cite{lan2021semantic}. As an illustration, we reference the AR architecture proposed in \cite{ren2019edge} that is divided into the user, edge, and cloud tiers (Fig. \ref{fig:ei}). The user tier senses the environment and transmits the raw video stream and other user controls to the edge tier. At the edge tier, image frames from the video stream are utilized to find a match with the cached images, for the retrieval of relevant information such as image annotations. If the image frame is not found from the cache, the frame is offloaded to the cloud for further matching. If a match is not found, computation is executed at the cloud and the edge cache is updated. Clearly, the image frames of the raw video streams are of heterogeneous importance. With AI-enabled semantic extraction and pre-processing of the video stream, the redundant transmission of repetitive or unimportant frames to the edge or cloud can be greatly reduced to alleviate the burden on backbone networks. Beyond semantic encoding for text, audio, or images, semantic communication has also emerged as a key enabler of efficient communications in distributed machine learning, e.g., gradient quantization schemes can significantly reduce the communication overhead of distributed AI model training.
            
           \textit{Edge resource optimization:} In a heterogeneous user network, it is of utmost importance that resources at the edge, e.g., for storage and computation, are well allocated to maximize the user QoE. AI-enabled solutions are increasingly utilized to solve the allocation problem given the dense distribution and mobility of users.  The study of \cite{guo2020adaptive} discusses that the rendering strategies of VR/AR users can be calibrated among local rendering, edge-assisted rendering, and edge-cloud rendering (i.e., local rendering of foreground interactions and edge rendering of background environment). The user QoE can be formulated as a function of latency and energy consumption, based on the user device and the required functions. Then, an effective rendering scheme can be formulated based on deep reinforcement learning (DRL) algorithm trained offline, subjected to the queue states at the edge servers and service requirements of the user. Moreover, the algorithm can be further refined using mechanism design when implemented online to account for the ad-hoc transitions in user usage requirements that may affect other users' QoE or rendering strategies. 
            
           \textit{Incentive mechanisms:} The stakeholders of the Metaverse, e.g., users, blockchain miners, and edge servers, each own valuable resources such as data and computation resources that can be leveraged for the enhancement of the Metaverse. To incentivize their participation, one may naturally consider a one-size-fits-all reward in which a homogeneous reward is allocated to all stakeholders. However, the result is that desirable stakeholders that can contribute more to the process, e.g., in terms of providing more resources for edge rendering, will lack the incentive to do so. As such, it is essential for the service requesters (e.g., VSP) to design incentive mechanisms to motivate the participation of these stakeholders. In light of the interactions among stakeholders and complex system states in the dynamic networks, AI approaches  have increasingly been proposed to design learning-based incentive mechanisms.

    \end{itemize}

\end{enumerate}

The edge intelligence empowered infrastructure layer connects all users in the Metaverse and supports its scalable, shardless, ubiquitous, and trustworthy realization.

\begin{figure*}
    \centering
    \includegraphics[clip, height= 6cm, width= \linewidth]{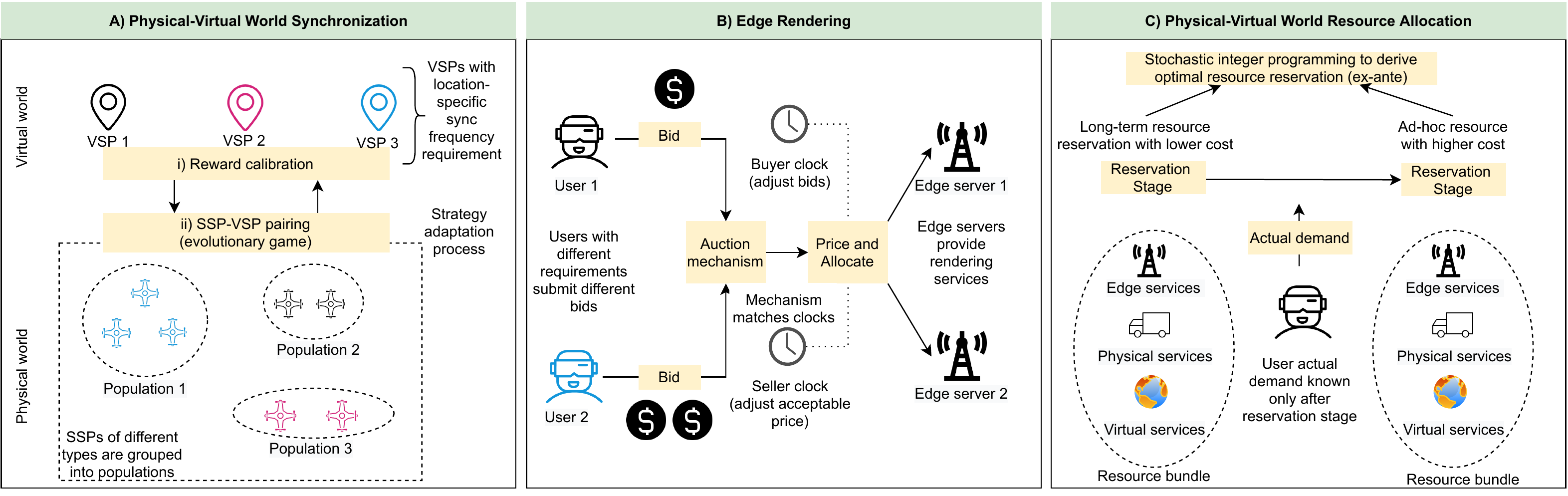}\par 
    \caption{We propose a framework for virtual city development in the Metaverse. In the first study, we propose collaborative sensing for the physical-virtual world synchronization. In the second study, we propose a pricing and allocation mechanism for edge rendering services among resource-constrained users. In the third study, we propose a resource allocation scheme that accounts for the unknown user demand to derive optimal resource reservation ex-ante.}
    \label{fig:seoul}
\end{figure*}

\subsection{Applications}

We identify some important emerging applications and services in Metaverse as follows. 

\subsubsection{Entertainment and social activities} Currently, entertainment and social activities are held on platforms that support audio and video transmission. Nevertheless, user  interactions are limited to rigid 2D grids of  users, and are still somewhat off what is experienced in the physical world. With the aid of VR and haptic technology, social interactions will be more immersive. 

\subsubsection{Pilot testing} Before products are being released in the market, they are usually tested by a small group of users in a controlled environment due to the cost of large-scale deployment or for safety reasons. The Metaverse will be a channel to pilot test products before they are released to the physical world at a low cost with fewer safety considerations. 

Moreover, users can have virtual twins of physical products delivered to their inventories directly in the Metaverse for marketing purposes. As an example, Hyundai has begun experimenting with providing virtual test drives for users albeit in the lower resolution Roblox world\footnote{https://www.roblox.com/games/7280776979/Hyundai-Mobility-Adventure}. In the Metaverse, test drive environments can be modeled exactly after highways with realistic traffic conditions. 

\subsubsection{Virtual education} The pandemic has necessitated the online delivery of education. However, a drawback of virtual education is the lack of personalization and difficulty of delivering ``hands-on" lessons. With more users in the Metaverse, the wealth of data can be used to further refine AI tutors for personalized lessons. Hands-on lessons that involve dealing with machines or tools can be delivered more effectively with haptics technology.

\subsubsection{Gig economy and creative industries} The Metaverse will mitigate the adverse effects of piracy on the gig economy and creative industry. The Metaverse will provide a platform for gig workers to create UGC and trade it actively as NFTs that uniquely identify the originality of the product, e.g., game object creation in GameFi\footnote{https://gamefi.org/}. When the product is transferred among buyers, a portion of the sales proceeds can be programmed to go to the creators automatically.

\section{Case Study: A Framework for Collaborative Edge-Driven Virtual City Development in the  Metaverse}

In this section, we present a case study of developing a virtual city in the Metaverse. For example, the development of ``Metaverse Seoul" has recently been proposed\footnote{https://www.euronews.com/next/2021/11/10/seoul-to-become-the-first-city-to-enter-the-metaverse-what-will-it-look-like} to cater to both tourists and local users, e.g., to access civil services online using HMDs. We motivate the collaborative edge-driven development of a virtual city in which the sensing, computation, communication, and storage resources at the network edge are leveraged to achieve the desirable qualities and features of the Metaverse (Fig. \ref{fig:seoul}). 

\subsection{Collaborative sensing for real-time physical-virtual world synchronization}

With continuous data synchronization, the virtual city is able to reflect the physical city in real-time. An enabling technology is collaborative sensing, in which IoT and wireless sensor networks are deployed to feed digital twins within the Metaverse with fresh data streams. 

In \cite{han2021dynamic}, we formulate a resource allocation problem in which SSPs (e.g., Drones-as-a-Service) are employed to collect data to maintain a regular sync between the physical and virtual worlds. The Unmanned Aerial Vehicle (UAV) fleets are owned by distinct SSPs, whereas the virtual city is maintained by distinct VSPs, each of which develops different areas of the virtual city that correspond to the real world. To employ the services of the SSPs, the VSP posts a reward pool (based on its budget) to be divided among SSPs that service the area. As more SSPs service the area, the data is uploaded at a higher frequency. However, each SSP receives a smaller proportion of the rewards and may churn to service other VSPs. To model the dynamic strategy adaptation of non-cooperative SSPs across the network, we utilize an evolutionary game based framework in which the SSPs are  clustered into populations based on their sensing capabilities, starting location, and energy cost. Using our evolutionary game based framework, we are able to model how the calibration of rewards by VSPs affect the composition of SSPs servicing it, and thereby simulate how the  synchronization frequency for each virtual region vary with the rewards provided.

\begin{figure}
    \centering
    \includegraphics[clip, height= 6cm, width= \linewidth]{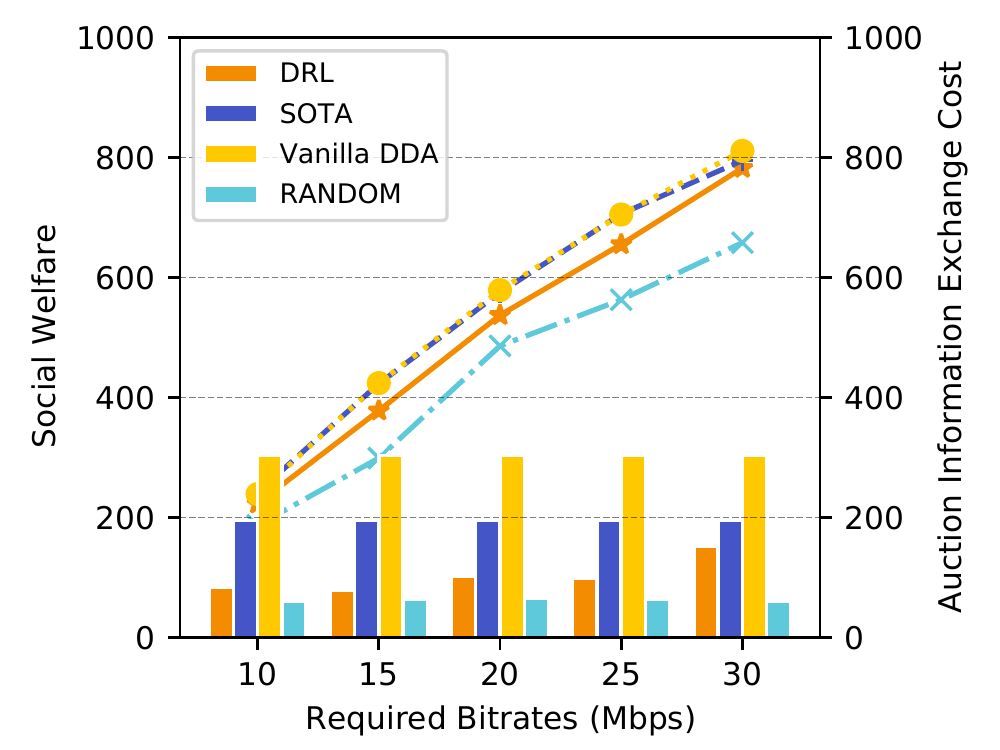}\par 
    \caption{In \cite{xu2021wireless}, we compare the DRL based DDA against the vanilla DDA and state-of-the-art method that adjusts the auction clock stepsize using the Ornstein-Uhlenbeck process \cite{friedman2018double}. The DRL based DDA can achieve  comparable social welfare (based on user QoE and edge server utility) at a much lower auction information exchange cost under various bitrates.}
    \label{fig:auction}
\end{figure}

\begin{figure}
    \centering
    \includegraphics[clip, height= 6cm, width= \linewidth]{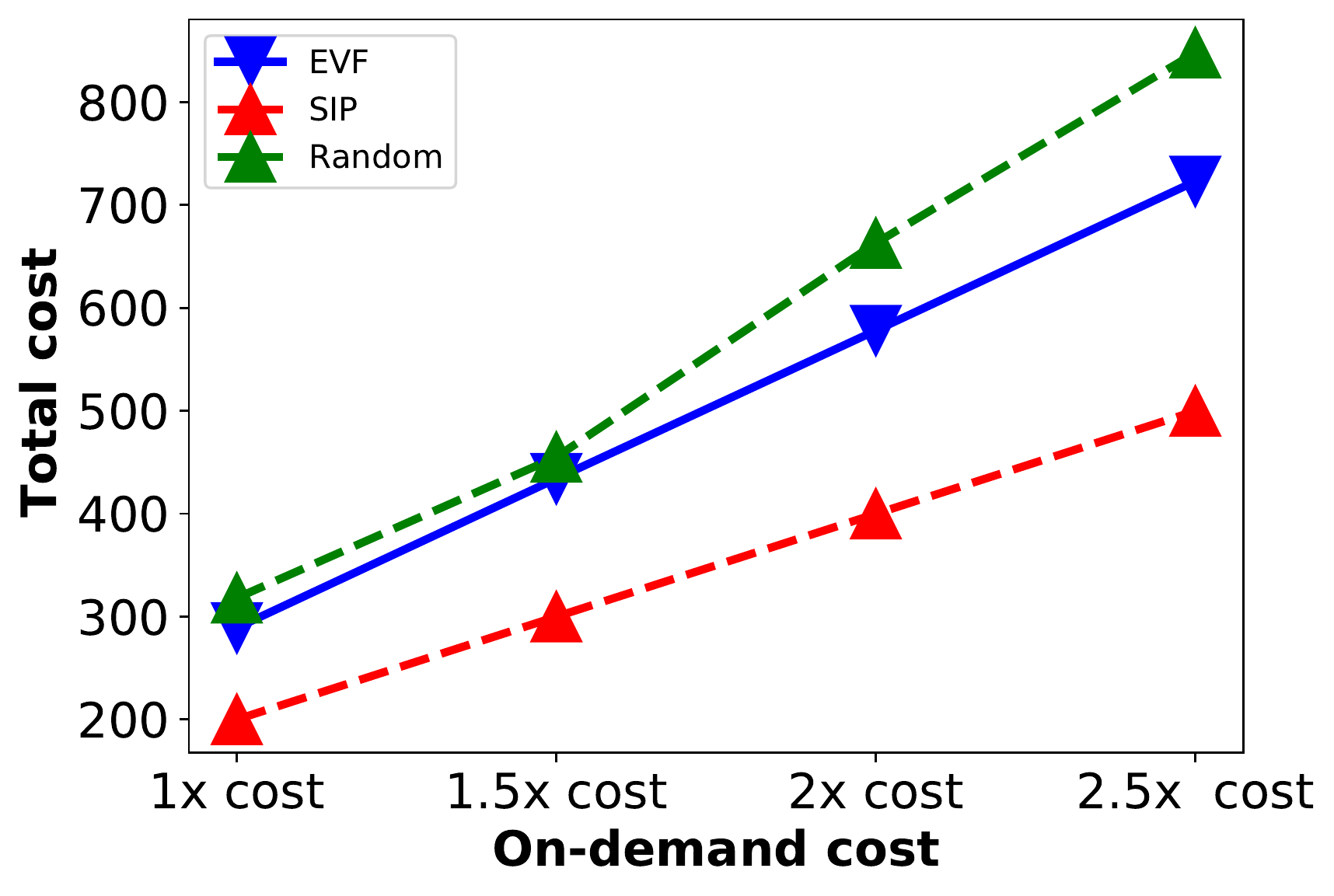}\par 
    \caption{In \cite{ng2021unified}, we compare the SIP with expected-value formulation (EVF) and the random scheme that models the user demand as the average historical value. The SIP can always achieve the best solution among the three to reduce the on-demand cost.}
    \label{fig:sip}
\end{figure}

\subsection{Edge-assisted efficient rendering of the immersive virtual world}

% Traditionally, panorama VR techniques render the 360-degree-view of the user. 

In light of battery limitations of user devices, non-panoramic VR rendering has been proposed such that only the images to cover the viewport of each eye are rendered, thereby demanding less data traffic and computation workload \cite{kelkkanen2020bitrate}. In \cite{xu2021wireless}, we study the provision of non-panoramic VR rendering services provided by edge servers and propose an incentive mechanism based on Double Dutch Auction (DDA)  for edge server-user association, as well as to price the services of edge rendering. The objective  is to allow VR rendering service providers to serve VR users in which their benefits (i.e., valuations of the services) are maximized. 

To derive the \textit{user valuation} of VR rendering services, we propose to formulate the user QoE as a function of Video Multi-Method Assessment Fusion (VMAF) and  Structural SIMilarity (SSIM) values. The former reflects the user’s perception of streaming quality, whereas the latter  measures VR image quality. The VMAF and SSIM values for a user in the Metaverse are in turn affected by the user's head rotation speeds (depending on VR functions) and expected bit rates of VR streaming from the edge. The \textit{edge server valuation} is formulated based on energy cost and the available computation and storage resources. To derive the edge server-user association, the users adjust their bids upwards, whereas the edge servers adjust their sell price downwards till a match in valuation is derived. The evaluation results show that the proposed incentive mechanism can motivate the providers and the users to participate rationally in the auction with desirable properties such as truthfulness. Moreover, we design a DRL-based auctioneer to accelerate this auction process by adjusting the stepsize of the auction clocks dynamically (Fig. \ref{fig:auction}).

\subsection{Resource allocation in the physical-virtual world ecosystem}

To support the Metaverse engine, VSPs have to leverage  both virtual and physical resources that are often owned by separate entities. For example, VSPs can utilize logistic services for physical goods delivery or edge services for computation offloading. 

Similar to other shared services (e.g., cloud services), such resources are usually priced based on two subscription plans i.e., reservation and on-demand plan. Generally, the reservation plan is cheaper than the on-demand plan, which is used on an ad-hoc basis when demand spikes. However, the VSP will need to decide on the resources to be allocated via the reservation plan before the \textit{actual} user demand is known (i.e., ex-ante). Therefore, a resource over-provisioning problem can occur if the VSP subscribes too many resources on the reservation plan. In contrast, a resource under-provisioning problem can happen if the VSP subscribes too little resources, i.e., the VSP has to use the more expensive on-demand plan. Taking into account the demand uncertainty of the users, we propose a two-stage stochastic integer programming (SIP) formulation for the VSPs in Metaverse to minimize its operation cost by allocating the resources across the two plans most strategically \cite{ng2021unified}. Using historical data on user demand, our  resource allocation scheme achieves a much lower cost than other schemes that do not consider the probability distribution of  user demand (Fig. \ref{fig:sip}). 

\section{Open Challenges and Future Research Directions}

\subsection{Redefining user QoE} 

The Internet has been optimized based on gradually evolving QoE metrics. Similarly, there exists a need to redefine the user QoE for the Metaverse. This requires interdisciplinary efforts, e.g., to draw relations among network requirements and user visual perceptions. For example, the human eye is unable to perceive images shown for less than  13 ms \cite{popovski2021internet}, thereby setting an upper-bound on the network timing requirements. Moreover,  VR applications in the Metaverse will place less emphasis on  the traditional focus of video resolution. Instead, foveated rendering  studies eye tracking to render important scenes and reduce the image quality of scenes in the peripheral vision \cite{patney2016towards}.

% DeepFovea utilizes generative adversarial networks (GAN) to model how our peripheral vision perceives the world, thereby reducing more than 90\% of the pixels required to be rendered for VR applications \cite{kaplanyan2019deepfovea}. 

\subsection{B5G and the Metaverse}

B5G communication systems will deviate from conventional metrics such as data transmission rate to Value of Information (VoI) \cite{popovski2021internet}, that accounts for both contents and age of the packet to be transmitted. As the Metaverse will feature novel and differentiated service provision, the supporting communication and networking infrastructure  must be semantic-aware and goal-oriented.

\subsection{Interoperability standards}

While tech companies race to compete for the upper-hand in the development of the Metaverse, the need to develop interoperability standards have arisen so that the vision for a seamless Metaverse can be realized. This is crucial to encourage the proliferation of UGC in the Metaverse. Moreover, a unified model to standardize the communication protocols of the Metaverse will eventually be necessary to enable access from diverse communication systems in different virtual worlds.  

\subsection{Security and Privacy}

The Metaverse will be built on blockchain-empowered economic ecosystems. As more transactions are conducted on the blockchain, the attack surface increases and  security concerns arise. For example, cyberattacks can utilize malicious smart contracts\footnote{\url{https://consensys.github.io/smart-contract-best-practices/known_attacks/}} to  gain access to the user's main crypto wallet. Moreover, new forms of hardware used to access the Metaverse bring about security challenges, e.g., the finger tracking of VR users can be used to infer the password.

In contrast to click-through rates for the Internet,  new dimensions of user data (e.g., eye tracking) can be collected and leveraged for more personalized advertising directly delivered to the FOV of users. This presents novel challenges to user data privacy.

\subsection{Economics of the edge-driven Metaverse}

The Metaverse will open up  novel possibilities of \textit{physical and virtual} service and resource trading among users and service providers. The contention for resources now extends from the physical to virtual world, in which rational users and service providers will have to optimize the resource usage efficiently in consideration of newly defined QoE.

\section{Conclusion }

In this article, we have discussed an architecture of the Metaverse and motivated the edge intelligence driven supporting infrastructure. Then, we present a case study of smart city development in the Metaverse, followed up with the future research directions. Our work serves as an initial attempt to motivate the confluence of edge intelligence and the Metaverse.

\bibliographystyle{IEEEtran}
\bibliography{fl-uav}

\section*{Biographies}
\small

{WEI YANG BRYAN LIM} is currently pursuing the Ph.D. degree (Alibaba Talent Programme) with the Alibaba-NTU Joint Research Institute (JRI), Nanyang Technological University (NTU), Singapore. His research interests include edge intelligence and resource allocation.

{ZEHUI XIONG}  is an Assistant Professor at Singapore University of Technology and Design. Prior to that, he was a researcher with Alibaba-NTU Joint Research Institute, Singapore. He received the Ph.D. degree in Computer Science and Engineering at Nanyang Technological University, Singapore. He was a visiting scholar with Princeton University and University of Waterloo. His research interests include wireless communications, network games and economics, blockchain, and edge intelligence.

{SUMEI SUN} [Fellow, IEEE] is the Principal Scientist, Acting Executive Director (Research), and the Head of the Communications and Networks Department with the Institute for Infocomm Research (I2R), Singapore. She is the Editor-in-Chief of IEEE Open Journal of Vehicular Technology, member of the IEEE Transactions on Wireless Communications Steering Committee, and a Distinguished Speaker of the IEEE Vehicular Technology Society 2018–2024. She's also the Director of IEEE Communications Society Asia Pacific Board and a member at large with the IEEE Communications Society.

{DUSIT NIYATO} [IEEE Fellow] is currently a Professor with the School of Computer Science and Engineering and, by courtesy, School of Physical and Mathematical Sciences, Nanyang Technological University, Singapore. He has published more than 380 technical papers in the area of wireless and mobile networking, and is an inventor of four U.S. and German patents. He was named the 2017–2021 Highly Cited Researcher in Computer Science. He is currently the Editor-in-Chief for IEEE Communications Surveys and Tutorials.

{XIANBIN CAO}   received the Ph.D. degree in signal and information processing from the University of Science and Technology of China, Hefei, China, in 1996. He is the Dean and a Professor with the School of Electronic and Information Engineering, Beihang University, Beijing, China. His research interests include intelligent transportation systems, airspace transportation management, and intelligent computation

{CHUNYAN MIAO}  is currently a professor in the School of Computer Science and Engineering, Nanyang Technological University (NTU), and the director of the Joint NTU-UBC Research Centre of Excellence in Active Living for the Elderly (LILY).

{QIANG YANG} [IEEE Fellow]  is the head of AI at WeBank (Chief AI Officer) and Chair Professor at the Computer Science and Engineering (CSE) Department of Hong Kong University of Science and Technology (HKUST).

\end{document}